\documentclass{elsart}

\usepackage{graphics}
\usepackage{graphicx}

\usepackage{amssymb}
\usepackage{amsmath}

\begin{document}

\begin{frontmatter}



\title{Development of a novel high-sensivitiy LAr purity monitor
based on an $\alpha$-source}


\author{A.~Badertscher}
\author{M.~Laffranchi,}
\author{A. Rubbia}
\address{Institut f\"{u}r Teilchenphysik, ETHZ, \\ CH-8093 Z\"{u}rich,
Switzerland}
\begin{abstract}
A novel liquid argon (LAr) purity monitor was developed with a
sensitivity to electronegative impurities of the order of ppb
(O$_2$ equivalent). Such a high purity is e.g. needed in a LAr
drift chamber. The principle is to measure the lifetime of
quasifree electrons in LAr, since this is the important parameter
for the operation of a drift chamber. Free electrons are produced
by ionizing the LAr with $\alpha$-particles emitted by the
$^{210}Po$ chain daughter of
an isotope $^{210}Pb$ source. From a measurement of the charge of the
electron cloud at the beginning and at the end of a drift path,
together with the drift time, the lifetime of the electrons is
obtained. The $\alpha$-particles have a very short range of about
50~$\mu$m in LAr and the ionization density is very high, typically
between $750\div 1500\rm\ MeV/cm$,  leading
to a high recombination rate. To suppress the recombination of the
argon ions with the electrons, the $\alpha$-source was put in a
strong electric field of $40\div 150$~kV/cm. This was achieved by
depositing the source on the surface of a spherical high voltage
cathode with a diameter of about 0.5~mm. The anode was also made
as a sphere of about the same diameter as the cathode, thus, close
to the axis between the two electrodes the electric drift field
was approximately a dipole field.
\end{abstract}

\end{frontmatter}

\section{Introduction} Liquid argon (LAr) detectors
need to monitor the purity of the LAr since electronegative
impurities (mainly $O_2$) capture ionization electrons, and hence
degrade the performance of the detector. Different types of liquid
argon purity monitors were developed for the LAr detectors in use
or for future experiments. For the LAr calorimeters in the H1
\cite{H1} and the Atlas \cite{Atlas} experiments the necessary
sensitivity of the monitors to electronegative impurities is of
the order of $ppm$ (oxygen equivalent). The pulse height spectra from $\beta$-decay
electrons and $\alpha$-particles are measured in a LAr ionization
chamber. In the drift chamber of the ICARUS detector \cite{Amerio:2004ze}
drift times of the order of $ms$ occur. In order to measure such
long drift times, it is necessary to purify the LAr from to a
level below $0.3\ ppb$ (oxygen equivalent). Purity monitors with this
sensitivity were built \cite{aqumon}, measuring the lifetime of
electrons which drift in
a homogeneous electric field over a distance of about 10~cm. The
drift electrons were extracted using an appropriately chosen photocathode,
which is flashed periodically with a bright light pulse.

Traditionally the problems encountered in designing purity
monitors were (1) related to the creation of a sufficiently large
drift electron cloud in order to produce clean signals
above noise and (2) to the extraction of the purity
with high precision and sensitivity.
 
The purity monitor described in this paper is also based on a
lifetime measurement of electrons. 
The method to determine the lifetime of electrons consists of
measuring the attenuation of the charge of an electron cloud
drifting in an electric field as a function of the drift time. The
mean lifetime of the electrons is obtained from equation
(\ref{exp}):
\begin{equation}
N(t_{drift}) = N_0 \cdot \exp{(-t_{drift}/\tau)}, \label{exp}
\end{equation}
where $N_0$ is the number of electrons at the beginning and
$N(t_{drift})$ the number of electrons at the end of the drift
path corresponding to a drift time $t_{drift}$.

However, our purity monitor includes
the following new features:
\begin{itemize}
\item a new almost monochromatic 
source of free electrons based on an energetic 5.3~MeV 
$\alpha$-source;
\item a dipole geometry to introduce a very high field
in the region of the cathode and anode, and a very low field in the
drift region in-between (inhomogeneous field);
\item a direct start and stop trigger for a source-event from the 
independent readout of the cathode and 
anode induced signals;
\item a built-in variation of the
drift time, due to the different path along the dipole
field lines introducing a spread-in-time for the arrival
of the electron cloud on the anode;
\item an event-by-event measurement of the drift
time and induced charges before the drift at the
cathode and after the drift at the anode, yielding
the attentuation as a function
of the event-by-event varying drift-time.
\end{itemize}

\begin{figure}[htb]
\begin{center}
\includegraphics[width=0.5\textwidth]{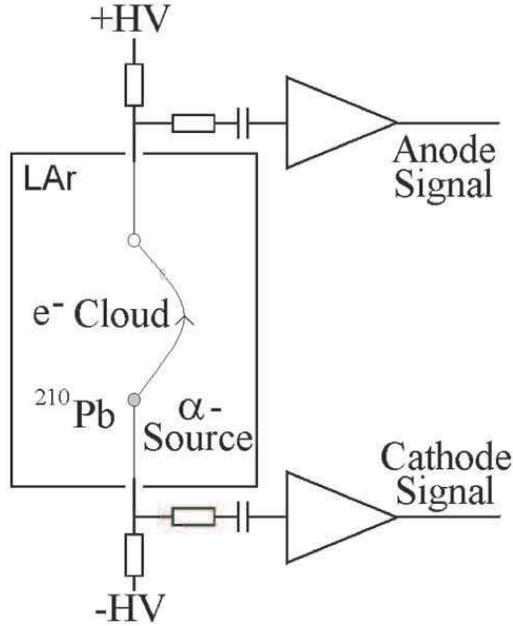}
\caption{\label{scheme} Schematic view of the purity monitor.}
\end{center}
\end{figure}

\begin{figure}[htb]
\begin{center}
\includegraphics[width=0.9\textwidth]{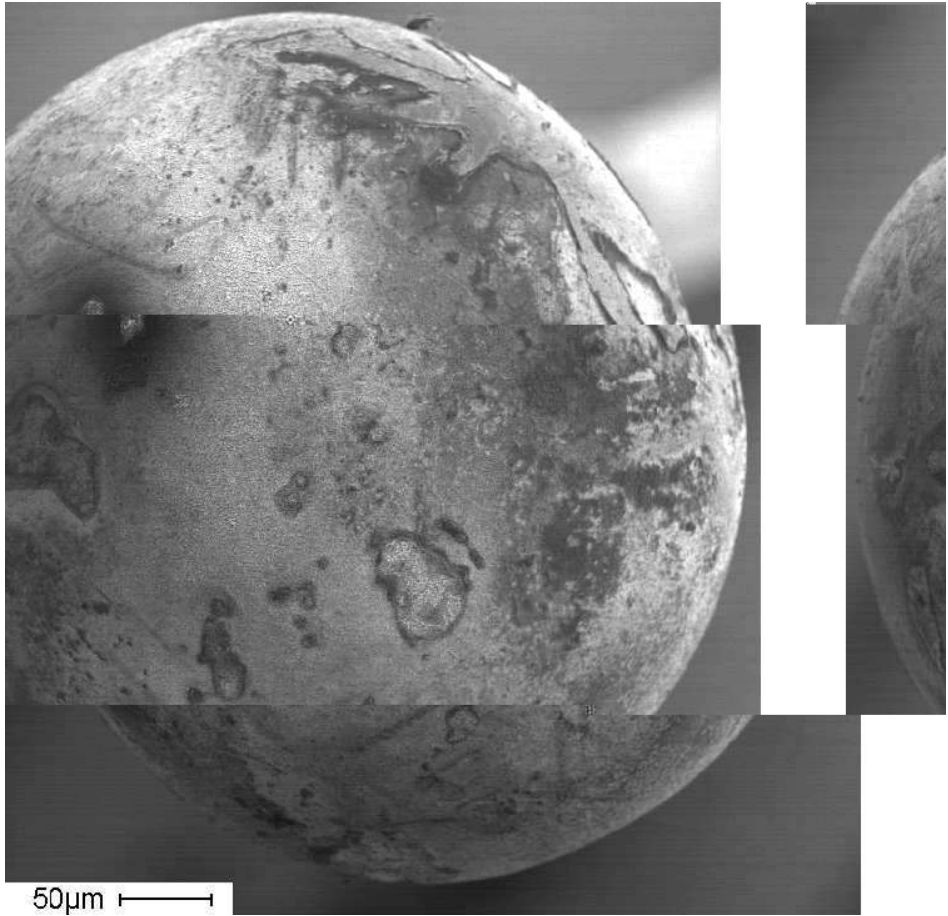}
\caption{\label{source} Electronic microscope
picture of the spherical platinum cathode with the
$^{210}Pb$ deposited on the surface. The scale is shown
in the lower-bottom corner.}
\end{center}
\end{figure}

\section{The purity monitor}
\label{sec:puri}
The purity monitor described in this paper is shown schematically
in Figure \ref{scheme}; it uses the ionization electrons produced
by the $5.3~MeV \alpha$-particles emitted by $^{210}Po$ to measure
the electron lifetime. The $\alpha$-emitter $^{210}Po$ is produced
with a decay fraction of almost 100\% through $\beta$-decays in
the decay chain of $^{210}Pb \rightarrow ^{210}Bi \rightarrow
^{210}Po$. The decay chain ends at the stable $^{206}Pb$ isotope.
In the decay chain of $^{210}Pb$ several $\alpha$- and
$\beta$-decays occur, but only the $\beta$-decay of $^{210}Bi$
with an endpoint energy of $1.2~MeV$ and the $\alpha$-decay of
$^{210}Po$ with an energy of $5.3~MeV$ have decay probabilities of
almost $100\%$, all other decays are very rare.

\begin{figure}[htb]
\begin{center}
\includegraphics[width=0.9\textwidth]{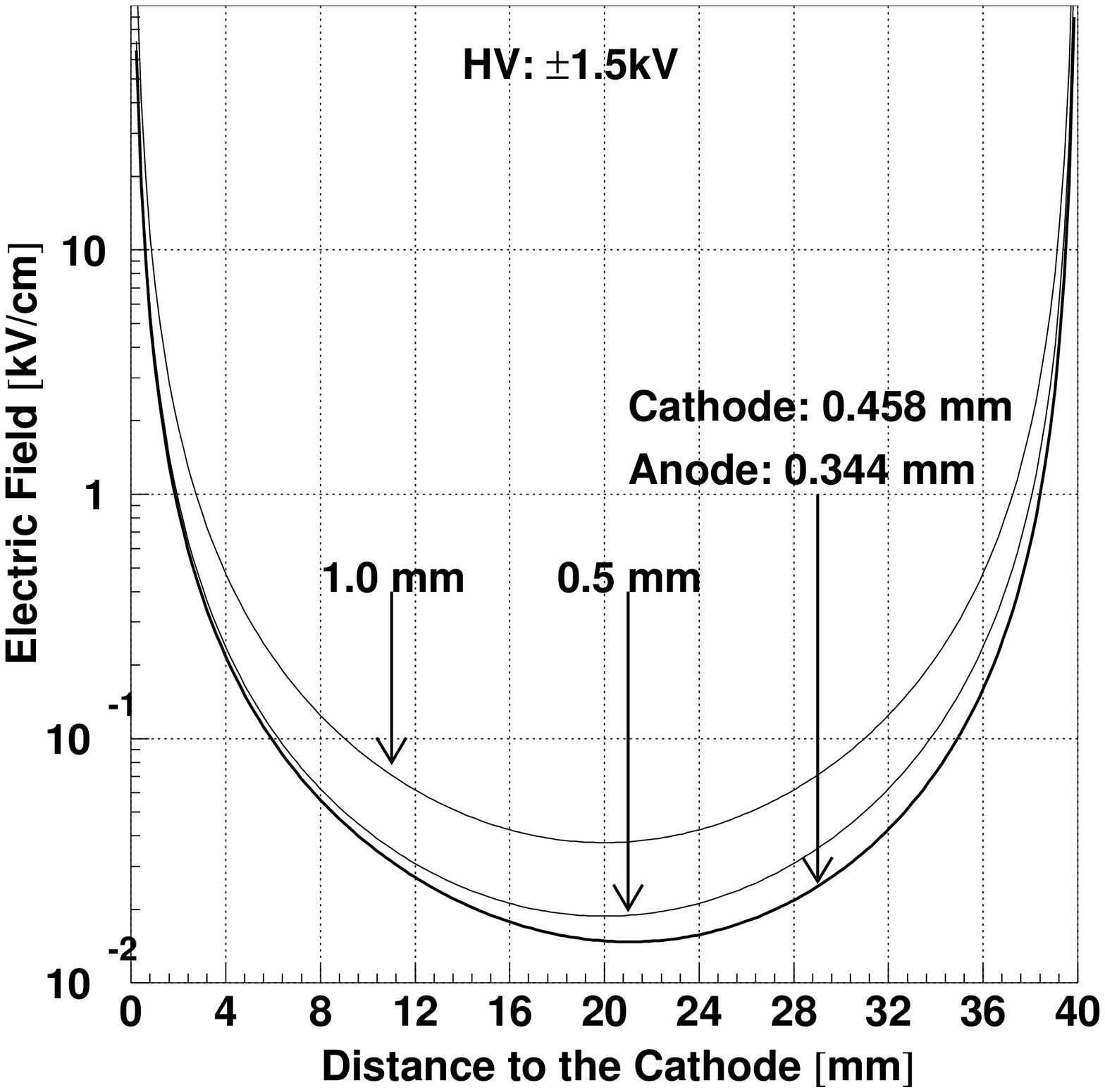}
\caption{\label{efield} Electric drift field on the axis between
the electrodes for different diameters of the electrodes and a
high voltage of $\pm$1.5~kV.}
\end{center}
\end{figure}

The high ionization density of $\alpha$-particles in LAr, typically
between $750\div 1500\rm\ MeV/cm$ as opposed to about 2~MeV/cm
for a m.i.p., leads to
an extremely high recombination rate \cite{Birks} of the argon ions with the
electrons along the track of the $\alpha$-particles, thus,
reducing the measurable electron charge \cite{diploma}. 
The recombination rate is reduced, when the ionization occurs in a
strong electric field \cite{Imel}. At a typical drift field of 500~V/cm, less 
than 1\% is recovered as free electrons. To expose the $\alpha$ source
to the mandatory electric fields of the order of $40\div 150$~kV/cm, it was
deposited onto the surface of a spherical platinum high voltage
cathode with a diameter of about 0.5~mm; applying a high voltage
of 2~kV produces an electric field $E\approx V/r$, where $r$ is 
the radius of the sphere, of about 80~kV/cm at the
surface. The spherical electrodes were made from a 76~$\mu$m thick
platinum wire by melting one end of the wire in a flame of a
Bunsen burner \cite{diploma}; the surface tension of the melting
platinum was forming spherical drops. The 20~kBq~$^{210}Pb$-source
\footnote{Purchased from AEA Technology QSA GmbH, D-38110
Braunschweig} with a mean lifetime of 31.9 years was dissolved  in
a 1.2 molar $HNO_3$-solution. A thin layer with an activity of
about $100~Bq$ was deposited electrolytically on the spherical
cathode. Figure \ref{source} shows an electron microscope picture
of a cathode, having a thin layer of $^{210}Pb$ deposited on the
surface; the diameter of the electrode is 458 $\pm 12~\mu$m.

\begin{figure}[htb]
\begin{center}
\includegraphics[width=0.9\textwidth]{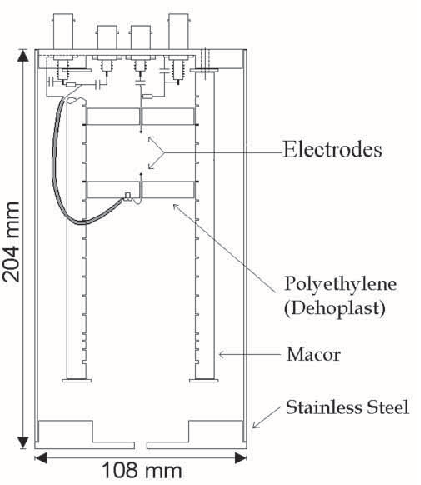}
\caption{\label{monitor} Mechanical design of the purity monitor.}
\end{center}
\end{figure}

The anode also consists of a Pt sphere with a diameter of 335 $\pm
7 ~\mu$m. A symmetrical high voltage $\pm HV$ applied to the two
electrodes produces approximately an electric dipole drift field.
The electric drift field on the axis between the electrodes is
shown in Figure \ref{efield} for a high voltage of 1.5~kV.

\begin{figure}[tb]
\begin{center}
\includegraphics[width=0.8\textwidth]{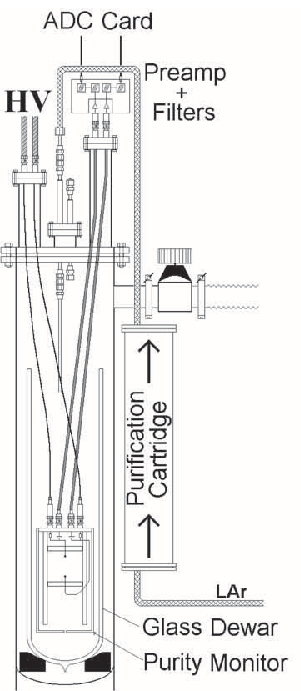}
\caption{\label{setup} Measuring set-up.}
\end{center}
\end{figure}

The number of free ionization electrons remaining after
recombination depends on the electric field along the entire track
of the $\alpha$-particle, i.e., it depends on the high voltage
applied to the cathode and its diameter. The range of the 5.3~MeV
$\alpha$-particles in LAr is only about 50~$\mu$m, i.e., 1/5 of
the used cathode radius. Thus, the electric field along the entire
track of the $\alpha$-particle can (in the worst case) decrease by
only 33\% from the value at the cathode surface. The number $N_0$
of electron-ion pairs produced by $\alpha$-particles depositing
their total energy of 5.3~MeV in LAr is $N_0 = 5.3 \cdot 10^6 eV/w
= 225 \cdot 10^3$, where $w = 23.6$~eV is the mean energy
needed to create an electron-ion pair in LAr. We stress
that this number is reduced if the alpha deposits a non-negligible
fraction of its energy in the lead (the range in lead is 16~$\mu$m). 
Neglecting this effect, we anticipate
here (see section~\ref{sec:edep} for more details) that the measured
quenching of this charge by the recombination (recombination
factor $R$) varied between 0.22 at a field on the cathode surface
of 44~kV/cm to 0.39 at 154~kV/cm (see Figure \ref{charge}), a variation
consistent with the Box model of recombination.

The charge of the electron cloud at the cathode and the anode is
obtained by integrating the current signals induced on the
electrodes by the movement of the electrons in the drift field.
The fast movement in the high field near the surface of the
electrodes induces a fast rising current signal (see Figure
\ref{pulses}) with a good signal to noise ratio. The argon ions
have drift velocities orders of magnitude smaller than the
electrons so that they do not contribute to the current signal. To
summarize, the configuration of Figure \ref{scheme} combines the
following desirable features:
\begin{itemize}
\item the high electric field on the cathode surface containing the
$\alpha$ source suppresses the recombination,
\item the fast drift velocity of the electrons in the high electric
field near the surface of the electrodes induces a short (approx.
1 $\mu$s) current pulse, which can be measured.
\item the small drift velocity of the electrons in the central
region of the dipole field (minimal field strength is a few V/cm)
allows to measure long drift times.
\end{itemize}


The mechanical design of the purity monitor is shown in Figure
\ref{monitor}. Two circular polyethylene plates with the
electrodes in the center are held by three Macor rods. The monitor
is shielded by a stainless steel cylinder and covered at the
bottom and at the top with steel plates. The distance between the
electrodes was varied between 20~mm and 50~mm. The measuring
set-up is shown in Figure \ref{setup}.

A glass dewar holding the purity monitor was mounted in a vacuum
chamber. Before filling the dewar with LAr, the chamber was heated
to about 70$^{\circ}$C for at least one day and pumped to a
pressure of about 10$^{-6}$~mbar. The LAr passed through a
5~$\ell$ purification cartridge containing 50\% BTS\footnote{Fluka
No. 18820, Fluka Chemie GmbH, CH-9471 Buchs SG, Switzerland}
catalyst and 50\% copper oxide. The BTS and the copper oxide were
reduced by controlled flowing of hydrogen gas through the cartridge
before it could be  used to purify LAr
from oxygen.

The readout electronics for the two electrodes is operated at room
temperature outside the vacuum chamber. It consists of a low noise
charge preamplifier of the type used for the ICARUS drift chamber
\cite{preamp} followed by a custom-made ac-coupled amplifier, which also acts
as a bandpass filter transmitting frequencies from 530~Hz to
760~kHz (-3 dB values). The preamplifier integrates the current
pulse from the electrode; its decay constant is about 250~$\mu$s.
The circuit has an overall sensitivity of 10.8~mV/k$e^-$
corresponding to 68~mV/fC. Both electrodes have their own readout
channels, which were carefully calibrated. The analog signals were
sampled with 10~MHz and digitized by a 12 bit ADC card in a PC;
the digital data were accumulated with a LabView program.

\section{Results}

\subsection{Signal shapes}
\begin{figure}[htb]
\begin{center}
\includegraphics[width=0.9\textwidth]{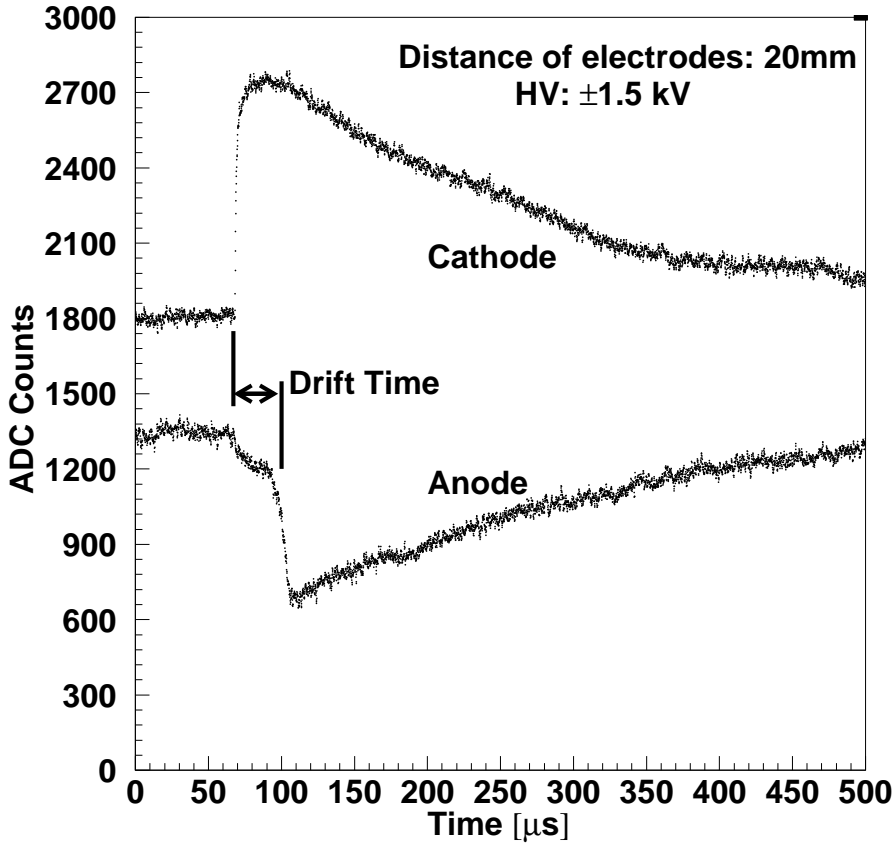}
\caption{\label{pulses} Measured cathode and anode pulse shapes
(raw data).}
\end{center}
\end{figure}

Figure \ref{pulses} shows the measured pulse shapes from the
cathode and the anode at a high voltage of $\pm$1.5~kV and an
electrode distance of 20~mm. The integrated current (i.e. the
charge) induced on the electrodes by the moving electron cloud is
shown as a function of time. The fast rising leading edge of the
signals is induced by the fast movement of the electrons in the
high field near the surface of the electrodes. The decay of the
signal is given by the decay time of the integrating electronics.
In the anode signal the contribution to the signal over the whole
drift time is seen: it starts with the fast movement of the
electron cloud near the cathode, gets flat during the time of the
slow drift through the central region of the drift field and rises
sharply when the cloud reaches the anode. In the cathode signal
the contribution from the drift in the low field region is hidden
in the rounding of the signal after the sharp rise. The charge is
given by the pulse height difference between the maximum (minimum)
and the base line defined by the (average) signal measured at
times before the cathode pulse starts.
\begin{figure}[htb]
\begin{center}
\includegraphics[width=0.9\textwidth]{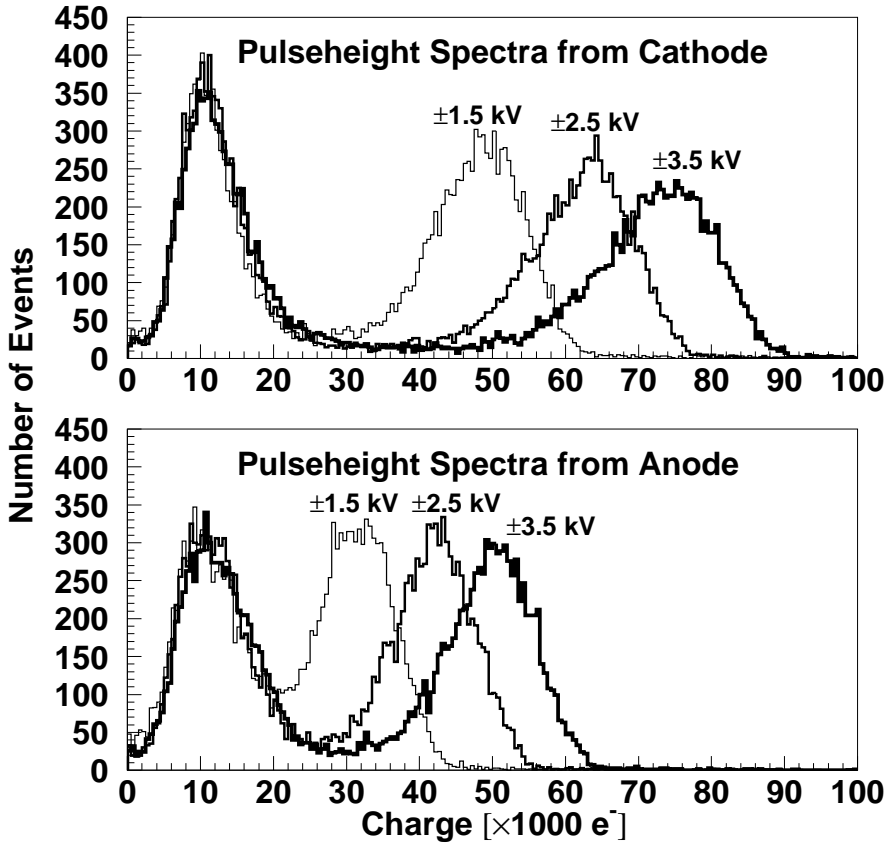}
\caption{\label{spectra} Measured cathode and anode pulse height
distributions. The anode was at a distance of 20~mm from the
cathode.}
\end{center}
\end{figure}
Figure \ref{spectra} shows the pulse height spectra from the
cathode and the anode measured at three different high voltages
and at a separation of the electrodes of 20~mm. The peaks from
$\alpha$-particles are clearly separated form noise and the small
pulses from $\beta$-decays ($\beta$-decay electrons have a maximal
range in LAr of 3~mm, compared to 50~$\mu$m for
$\alpha$-particles, and hence, have a different distribution of
induced signals between cathode and anode). The energy deposited
by the $\alpha$-particles in LAr is reduced from the maximal
energy of 5.3~MeV and smeared out by the energy deposition in the
lead of the source (the range in lead is 16~$\mu$m). The maximal
charge (for a fixed high voltage) is given by the end point of a
pulse height spectrum and corresponds to the deposition of the
total energy of 5.3~MeV in the LAr. The values for the maximal and
the mean (corresponding to the peak of the distribution) measured
charge are given in Table~\ref{tab1}. 

\begin{table}[ht]
\begin{center}
\begin{tabular}{|l|c|c|c|c|}\hline
HV ($kV$)  & Average charge  &Maximal charge& Minimal drift time & Lifetime $\tau$       \\
           & [$10^3$ electrons] & [$10^3$ electrons] & $[\mu s]$  & $[\mu
           s]$ \\
\hline
1.0           &36 &50  & 20 & 83 $\pm$2             \\
1.5           &49 &62  & 14  & 105$\pm$2            \\
2.0           &58 &70  & 12.5  & 112$\pm$2          \\
2.5           &62 &76  & 11   & 109$\pm$2           \\
3.0           &69 &82  & 10.5   & 120$\pm$3         \\
3.5           &72 &88  & 10   & 108$\pm$3           \\
\hline
\end{tabular}
\end{center}
\caption{ \label{tab1} The average and the maximal electron charge
measured at the cathode, the minimal drift time and the lifetime
measured at different high voltages for an electrode distance of
20~mm are given. The errors given for the lifetime are statistical
fit errors only.}
\end{table}

\subsection{E-field dependence}
\label{sec:edep}
\begin{figure}[htb]
\begin{center}
\includegraphics[width=0.9\textwidth]{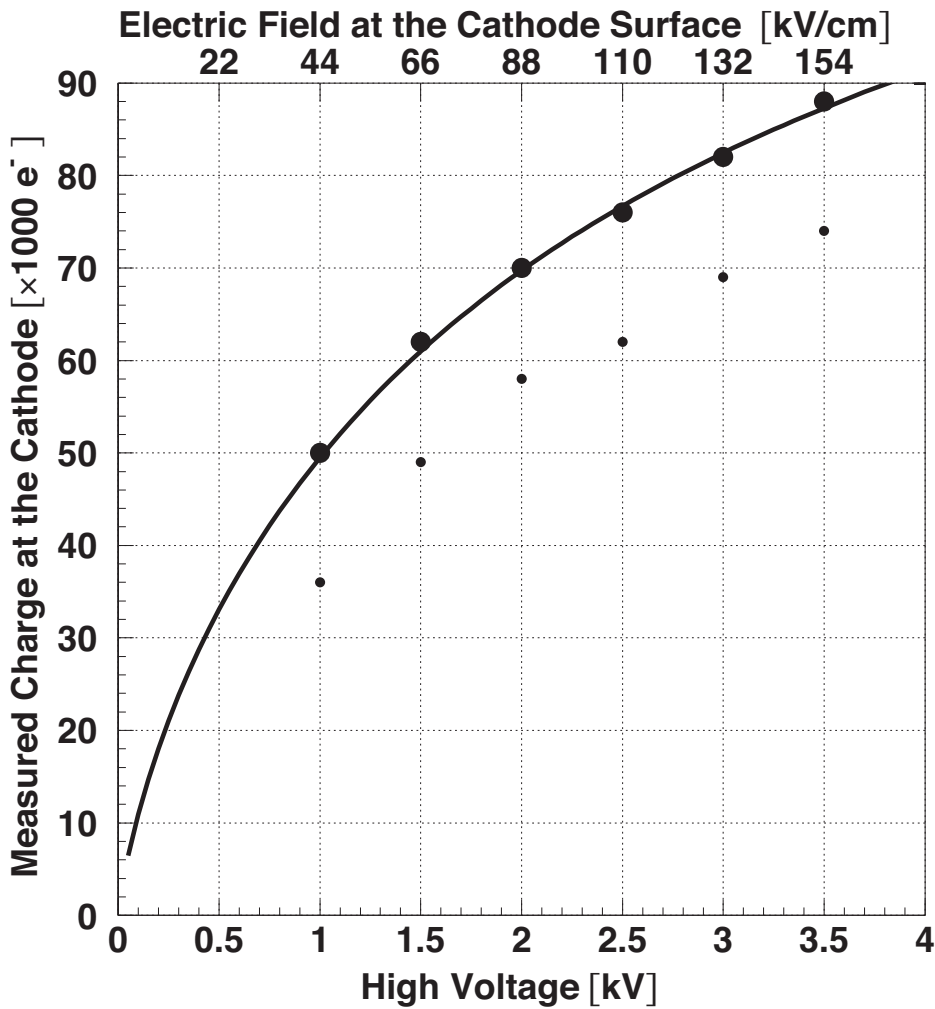}
\caption{\label{charge} Maximum measured charge at the cathode
as a function of
the applied high voltage (or as a function of the electric field
on the cathode surface). Also shown is the average measured charge
(see text for details).}
\end{center}
\end{figure}

In order to study the recombination of the electron-ion
pairs produced by $\alpha$'s as
a function of the electric field, we consider
the maximal charge (i.e. the end-point) of the electron clouds
measured at the cathode. In this way, we try to suppress
the uncertainties related to the energy loss of the $\alpha$'s
in the source Pb before they enter the liquid Argon medium\footnote{We
note that regardless of the actual thickness of the deposit of
lead on the Pt sphere, solid angle considerations limit the statistics
of $\alpha$'s to the few outer microns thickness.}.
In Figure \ref{charge} the maximum measured charged
is plotted as a function of the high voltage (or as a function
of the electric field on the cathode surface). The strong
dependence on the electric field supports 
the interpretation of the events in favour of the $\alpha$-decays;
the $\beta$-decays would never show such a dependence.
To our knowledge, this curve represents the
first measurement of the recombination factor for
$\alpha$-particles in LAr as a function of the electric field at
field strengths $\gtrsim 40\rm\ kV/cm$.

The observed curve fits well with
the Box Model\cite{Imel}:
\begin{equation}
\frac{N}{N_0} = \frac{E}{C} \ln(1 + \frac{C}{E}),
\end{equation}
where $N$ is the number of electrons after recombination and $N_0$
before the recombination. $C$ is a constant depending on the
ionizing particle and the medium, but not on the field. For the field,
we use $E=f\times V$ where $V$ is the potential of the cathode,
and $f$ is the amplification due to the sphere geometry. For a sphere
of radius $r$, one expects $f=1/r$.
From a fit to the measured curve as a function
of the cathode voltage, we can extract $C=214\rm\ cm/kV$,
$f=42\ \rm cm^{-1}$ and $N_0 = 141\cdot 10^3$ electrons. The 68\% C.L.
for the parameters assuming a 10\% uncorrelated error on the points
is $159<C<299\rm\ cm/kV$, $29<f<55\ \rm cm^{-1}$ and $(112<N_0<198)
\cdot 10^3$ electrons. A systematic variation of the measured points
by $\pm 10\%$ does not change appreciably the fitted $C$ and $f$ parameters
which depend mostly on the changing slope of the curve.

The fitted amplification factor $f=42\ \rm cm^{-1}$ is in perfect agreement
with the measured properties of the cathode, since the
radius of a sphere corresponding to such an amplification
is $r\approx 1/f=240\ \mu m$\footnote{We recall here that
the range of the $\alpha$ being so short, the error introduced
by the variation of the field over the range is less than 30\%.}. 
The measured
diameter of the electrode (see section~\ref{sec:puri}) 
is 458 $\pm 12~\mu$m.

In order to compare this result to expectations,
we developed a simulation of the $\alpha$-source.
The number of ionization
electrons was obtained by numerically integrating $dN/dx$ of the
semi-empirical expression of Birks \cite{Birks} along the track of
the $\alpha$-particles \cite{diploma}:
\begin{equation}
\frac{dN}{dx} = \frac{\frac{dE}{dx} \frac{1}{w}}{1 +k_B(E)
\frac{dE}{dx}},
\end{equation}
where $w$ is the mean energy to produce an electron-ion pair in
LAr, $w$ = 23.6~eV. 
The Birks law parameters were extracted from measurements in a
LAr TPC with m.i.p. or stopping protons up to 
$\approx 30\rm\ MeV/cm$ at electric fields up
to 500~V/cm\cite{Cennini:ha}. In order to predict the
recombination of the $\alpha$, one must extrapolate
these parameters to  very high ionization densities,
between 750 and $1500\rm\ MeV/cm$.
The E-field dependence of the Birks factor
$k_B(E)$ was obtained from the Box Model. 
For $\alpha$-particles in LAr, we obtain $C = 210\rm\ cm/kV$. This
is in excellent agreement with the observed shape of the curve.

\subsection{Drift electron lifetime}
\begin{figure}[htb]
\begin{center}
\includegraphics[width=0.9\textwidth]{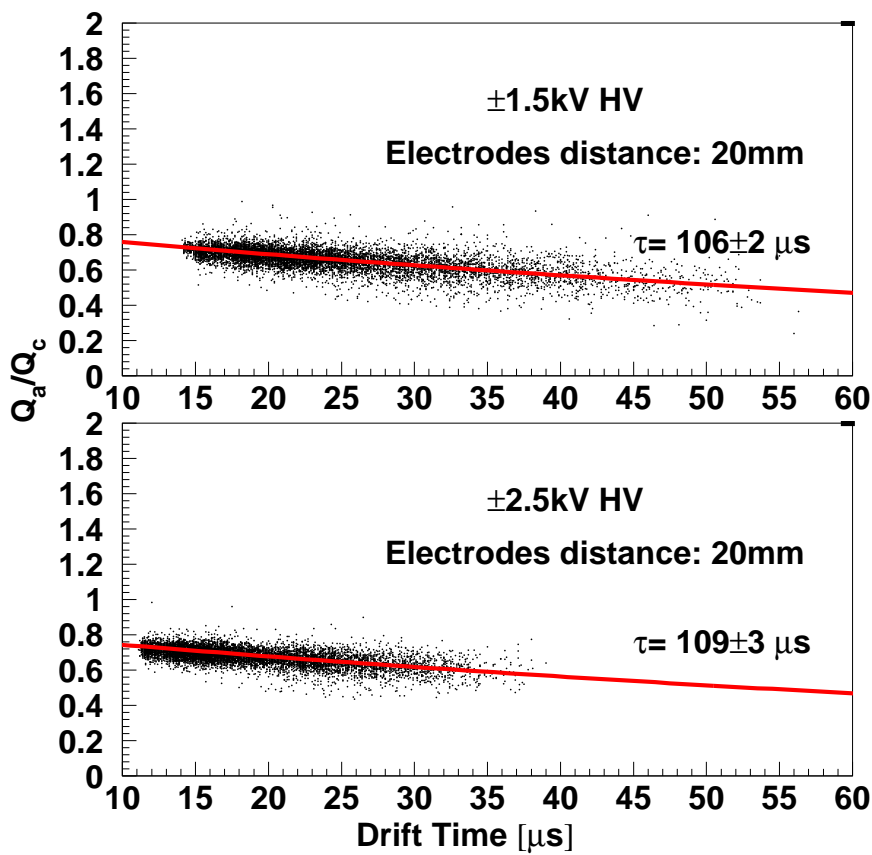}
\caption{\label{tau0} The anode to cathode charge ratio is plotted
as a function of the drift time. From an exponential fit the mean
lifetime $\tau$ is obtained.}
\end{center}
\end{figure}

Selecting $\alpha$-particle events with a cut on the pulse height,
the measured anode to cathode charge ratio $Q_a/Q_c$ for each
selected event is plotted versus the drift time in Figure
\ref{tau0}, for a distance of the electrodes of 20~mm and high
voltages of 1.5 and 2.5~kV. The drift time of the events
varies between $15\div 55 \mu s$ for the 1.5~kV and
 $12\div 35 \mu s$ for the 2.5~kV. 
This variation is due to different drift paths of the electrons in the dipole
field: the minimal drift time corresponds to electrons drifting on
the axis between the electrodes; electrons starting off axis on
the cathode sphere follow the dipole field lines, i.e., they have
a longer drift path and, in addition, feel the smaller drift field
in the central region. A direct fit of an exponential decay function to
the charge ratio as a function of the drift time yields the mean
lifetime $\tau$ of the electrons. Table \ref{tab1} summarizes the
results obtained with different high voltages for an electrode
distance of 20~mm.

The capture cross section of electrons by electronegative
impurities depends on the electric drift field \cite{capture},
i.e., the observed drift-electron lifetime can also depend on the 
drift field and not only
on the concentration of impurities. Since the drift field in our
purity monitor is very inhomogeneous, this effect can in principle
also distort the exponential decay curve of the charge. However,
for drift fields $\leq$ 600~V/cm the lifetime of electrons in LAr
for a $O_2$ concentration of 3.5~ppb was measured to be almost
constant \cite{edep}. As seen from Figure \ref{efield}, the drift
path in the high field ($\geq$ 1~kV/cm) is only a very small
fraction of the total drift path and contributes very little to
the drift time. Thus, only a small correction, depending on the
applied high voltage and the distance between the electrodes, has
to be applied to the measured drift time to normalize it to a
constant drift field. This correction was not applied to
the lifetimes given here. 

\section{Conclusion}
To conclude, we have developed a novel LAr purity monitor using an
$\alpha$-source in a very high electric field to produce the free
drift electrons. The adopted dipole geometry has allowed to
avoid the otherwise typical strong quenching of the $\alpha$.
We have measured the recombination factor of the
ionization charge from $\alpha$-particles as
the function of the electric field, ranging from
$40\div 150$~kV/cm.

In a series of measurements performed
in a dedicated setup, drift electron lifetimes of the order of 100~$\mu$s
were measured at electrode distances of 20~mm with a precision of
2-5 \%. To measure longer lifetimes, up to a few ms, larger
electrode distances could be needed, making it necessary to
install field shaping electrodes in the long drift region. 

\section*{Acknowledgements}
We thank Prof. B. Eichler from the Radiochemistry Department of
the Paul Scherrer Institut (PSI), CH-5232 Villigen PSI, for his
advice and for his help to prepare the $\alpha$ source.
We thank P.~Picchi, F.~Pietropaolo and F.~Sergiampietri
for useful discussions and suggestions. This work was supported
by the Swiss National Research Foundation.


\end{document}